\documentclass{article}
\usepackage{graphicx} 
\usepackage[
 left=2.5cm,
 right=2.5cm,
 top=2.5cm,
bottom=2.5cm,]{geometry}
\usepackage{siunitx}
\usepackage{amsmath}
\usepackage[hidelinks]{hyperref}
\usepackage{booktabs}
\usepackage{float}

\usepackage{makecell}
\usepackage{xurl}
\usepackage{lineno}

\usepackage{setspace}

\usepackage{authblk}

\DeclareSIUnit\h{h}
\DeclareSIUnit\bpm{bpm}
\DeclareSIUnit\min{min}

\usepackage[backend=biber,style=nature]{biblatex}
\title{\textbf{Continuous non-contact vital sign monitoring of neonates in intensive care units using RGB-D cameras}}

\author[1,*]{Silas {Ruhrberg Estévez}}
\author[1]{Alex Grafton}
\author[2]{Lynn Thomson}
\author[1]{Joana Warnecke}
\author[2,3]{Kathryn Beardsall}
\author[1]{Joan Lasenby}

\affil[1]{Department of Engineering, University of Cambridge, Cambridge, UK}
\affil[2]{Rosie Hospital, Cambridge University Hospitals NHS Foundation Trust, Cambridge, UK}
\affil[3]{Department of Paediatrics, University of Cambridge, Cambridge, UK}
\affil[*]{Corresponding author: Silas {Ruhrberg Estévez}, sr933@cam.ac.uk}
\date{}

\begin{document}

\maketitle

\begin{abstract}
Neonates in intensive care require continuous monitoring. Current measurement devices are limited for long-term use due to the fragility of newborn skin and the interference of wires with medical care and parental interactions. Camera-based vital sign monitoring has the potential to address these limitations and has become of considerable interest in recent years due to the absence of physical contact between the recording equipment and the neonates, as well as the introduction of low-cost devices. We present a novel system to capture vital signs while offering clinical insights beyond current technologies using a single RGB-D camera.  Heart rate and oxygen saturation were measured using colour and infrared signals with mean average errors (MAE) of $\SI{7.69}{\bpm}$ and $3.37\%$, respectively.  Using the depth signals, an MAE of $4.83$ breaths per minute was achieved for respiratory rate. Tidal volume measurements were obtained with a MAE of $\SI{0.61}{\ml}$. Flow-volume loops can also be calculated from camera data, which have applications in respiratory disease diagnosis. Our system demonstrates promising capabilities for neonatal monitoring, augmenting current clinical recording techniques to potentially improve outcomes for neonates.
  
\end{abstract}
\section{Introduction}
Each year there are approximately $135$ million live births globally~\cite{Births}. It is estimated that around $3-10\%$ of neonates, depending on their gestational age,  receive some level of care on a neonatal intensive care unit (NICU)~\cite{Kim2021},~\cite{Talisman2023}. Common reasons for admission to a NICU include respiratory distress syndrome, bradycardia, and infection~\cite{NICUadmissionreasons}. Preterm birth, defined as birth occurring before $37$ weeks of gestation, is a significant risk factor for NICU admission. It occurs in up to $10\%$ of all births and is responsible for approximately $1$ million neonatal deaths worldwide each year~\cite{Births}~\cite{Blencowe2013}. Monitoring on the NICU can be achieved both using biosensors and examinations by healthcare professionals. For detailed evaluations, quantitative assessments of vital signs are often performed~\cite{Kumar2019}. \\

 Vital signs reflect changes in physiological functions and standard clinical care in NICUs involves monitoring of heart rate, respiratory rate, body temperature, blood glucose levels, and oxygen saturation~\cite{vitalsigns}.  Notably, bradycardia, characterised by a significantly reduced heart rate, is predictive of sepsis~\cite{Ohlin2010}, while prolonged hyperventilation has been associated with adverse clinical outcomes~\cite{Bifano1998},~\cite{Stenzel2020}. Abnormal respiratory rates are also linked to increased risks of cardiac arrest, metabolic acidosis, and blood gas imbalances~\cite{Cretikos2008}. Normal oxygen saturation target values for newborns range from $90\%$ to $95\%$~\cite{Kayton2017}. Values higher than this range are harmful, particularly in preterm infants, as they can lead to increased generation of oxygen free radicals and ischaemia–reperfusion injury~\cite{Saugstad2010}. Lower saturation levels indicate a need for oxygen supplementation to ensure adequate oxygen delivery to tissues. Continuous monitoring of vital signs allows for early detection of complications and timely intervention to improve neonatal outcomes in the NICU setting~\cite{Kumar2019}.\\

Respiratory function can be characterised by various metrics, ranging from simple respiratory rate to  sophisticated assessments using spirometry, which measures airflow dynamics during breathing cycles to quantify volume and flow changes~\cite{Moore2012}. In clinical settings,  rapid assessment of neonatal breathing can be achieved using the Silverman-Anderson~\cite{Silverman}  and Downes' score~\cite{Downes}. However, both methods require clinicians to observe the patients and are thus subject to individual biases. During intensive care unit stays, thoracic impedance is often used to measure respiratory rate. This approach involves passing a small electrical current through surface electrodes placed on the chest and measuring the resulting voltage changes, which correlate with chest expansion during breathing~\cite{Wilkinson2009}. However, non-respiratory movements can interfere with the measurement by altering the voltage, leading to contamination of the respiratory rate data at an amplitude much larger than the respiratory signal of interest. In adults, a common method for monitoring respiratory function involves constructing flow-volume curves through maximum effort spirometry~\cite{Sterner2009},~\cite{Lizal2018}. However, this is not a feasible monitoring option for neonates who cannot perform maximum effort breaths on command. Instead, tidal volumes from normal breaths serve as an approximate metric~\cite{Schmalisch2005}. \\

Neonatal skin is exceptionally delicate, and prolonged electrode attachment can lead to iatrogenic skin injuries~\cite{Chung2019}. Furthermore, the presence of monitoring equipment may impede routine clinical care and hinder parent-infant interactions~\cite{Chung2019}. Balancing  accurate monitoring and potential risks to neonatal skin integrity and parent-child bonding remains a challenge in neonatal intensive care settings. Various sensing devices necessitate separate attachments, each presenting unique limitations. Heart rate is commonly measured using either an ECG or a pulse oximeter. Although ECGs offer higher accuracy, they require attaching electrodes to the skin for extended periods. These electrodes can also be used for  thoracic impedance measurements.  Oxygen saturation is estimated from a clip-on pulse oximeter that records the absorbance of light of different wavelengths which depends on the amount of oxygenated haemoglobin present in arterial blood. \\

In the past 15 years, various non-contact approaches have been explored for measuring vital signs in NICUs.  RGB cameras have been utilised for heart rate monitoring~\cite{Villarroel2019},~\cite{Chen2020},~\cite{Paul2020},~\cite{Svoboda2024} and oxygen saturation~\cite{Ye2024}. RGB-D cameras have been trialled for respiratory rate and volume measurements~\cite{Yu2012},~\cite{Kyrollos2021}. RGB-D cameras capture video streams containing colour images (red, green, and blue), infrared images, and depth images, the latter of which measures the distance of objects in the scene from the camera for each pixel using time-of-flight technology. The increased usage of cameras is facilitated by recent improvements in image quality for portable low-cost solutions.  Despite promising results from non-contact methods, comprehensive frameworks for monitoring all vital signs simultaneously remain elusive in clinical settings. Many studies have been limited by small sample sizes, the availability of accurate ground truth signals, or were conducted under controlled conditions that may not fully represent unaltered and real-world NICU environments. Further research is needed to address these limitations and develop non-contact monitoring systems capable of accurately and reliably tracking all vital signs in neonatal intensive care settings. \\

We conducted a clinical study at the Rosie Hospital in Cambridge to evaluate the feasibility of using continuous non-contact monitoring for infants in a real-world NICU setting. The primary objective of this study is to determine whether RGB-D cameras can accurately measure vital signs, such as heart rate, respiratory rate, and oxygen saturation, which are typically recorded using sensors attached to the skin of the neonates. Importantly, our method achieves this without physical contact or modifications to the clinical environment. Additionally, we aim to measure other critical parameters, including tidal volume and flow-volume dynamics, which cannot be recorded with standard NICU equipment. Furthermore, we will provide a publicly available dataset of the pose estimations with the resulting depth and colour signals along with the measured ground truth data and the analysis code to facilitate the development of new algorithms in this field.\\

\begin{figure}[H]
    \centering
    \includegraphics[width=\textwidth]{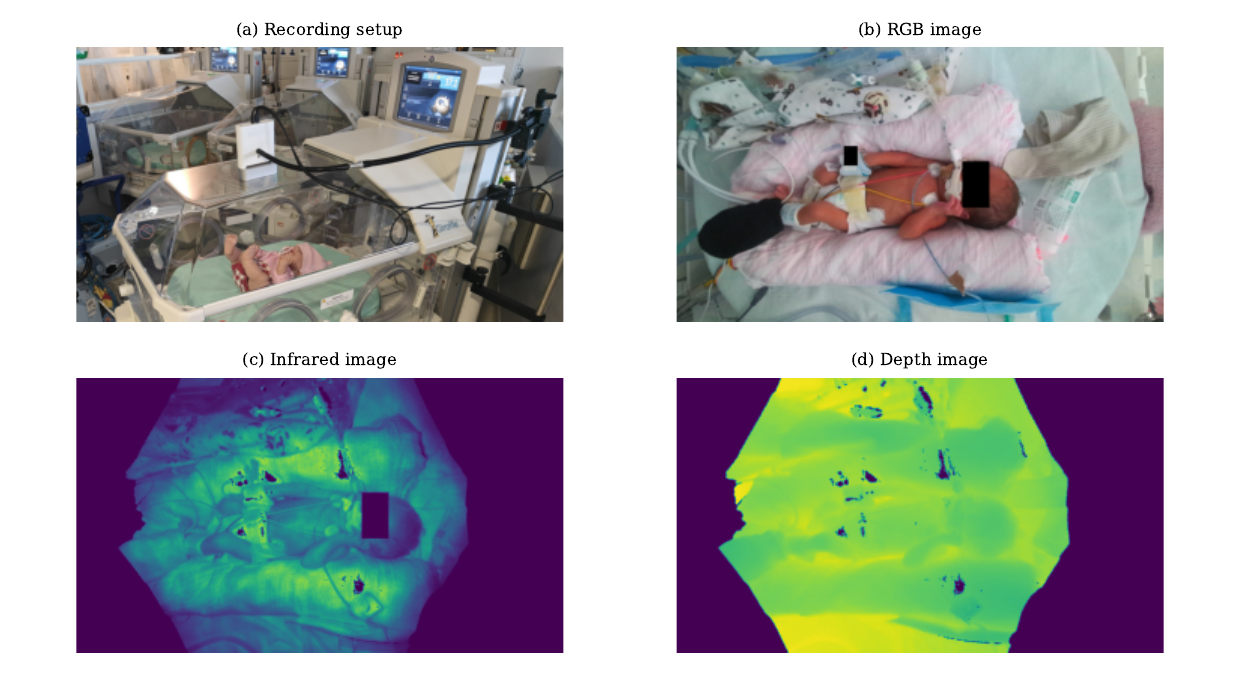}
    \caption{\textbf{Recording setup used in the clinical study and resulting images.} (a) Mounting of the Azure Kinect camera to the incubator using a flexible arm.  A doll is used for visualisation purposes. (b) RGB image recorded by the Azure Kinect camera. (c) Infrared image of the same neonate. (d) Depth image derived from the infrared image using time-of-flight. In the RGB and infrared images, the facial region of the baby and other identifiable objects are obscured by a black square. The pixel values of the infrared and depth images have been scaled for visualisation purposes, where darker regions correspond to distances closer to the camera. Black pixels in both the infrared and depth images indicate areas where insufficient infrared reflection occurred. Additionally, a phototherapy mask is applied to the left foot to shield the pulse oximeter from infrared radiation.}
    \label{fig:setup}
\end{figure}

To achieve these objectives, we utilised a single RGB-D camera (Microsoft Azure Kinect) mounted on the incubator in a manner that did not disrupt clinical procedures (see Fig. \ref{fig:setup}). Ground truth signals from standard NICU equipment were employed to validate the vital signs derived from the camera data. For respiratory monitoring, infants receiving mechanical ventilation were selected as ground truth due to the recognised inaccuracies associated with thoracic impedance measurements~\cite{Villarroel2019}. \\

\section{Results}
Data were collected from August 2021 to May 2024 as part of a larger collaboration between the University of Cambridge Engineering Department and the Rosie Hospital NICU. A total of 14 preterm infants were included in the vital sign monitoring study.  The study population consisted of 11 males and 3 females, with 8 infants from white backgrounds and 6 from non-white backgrounds. Respiratory ground truth measurements from ventilators were available for 3 neonates. Each neonate was recorded for approximately $\SI{1}{\h}$ with the Azure Kinect RBG-D camera while simultaneously measuring vital signs using the standard equipment available on the NICU. A summary of patient demographics is provided in Table \ref{tab:patient_demographics}.\\

\begin{figure}[!htb]
    \centering
    \includegraphics[width=\textwidth]{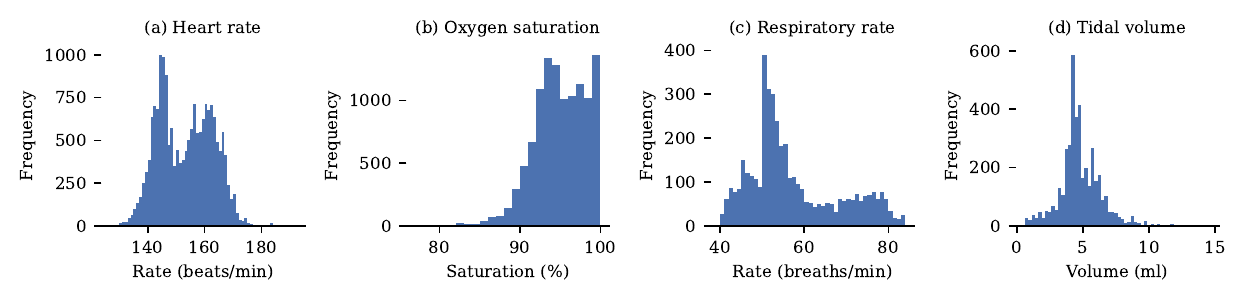}
    \caption{\textbf{Characterisation of ground truth data set.} (a) Heart rate from ECG, (b) oxygen saturation from the pulse oximeter, (c) respiratory rate from ventilator, and  (d) tidal volume from ventilator. }
    \label{fig:groundtruthdata}
\end{figure}

\begin{figure}[!htb]
    \centering
    \includegraphics[width=\textwidth]{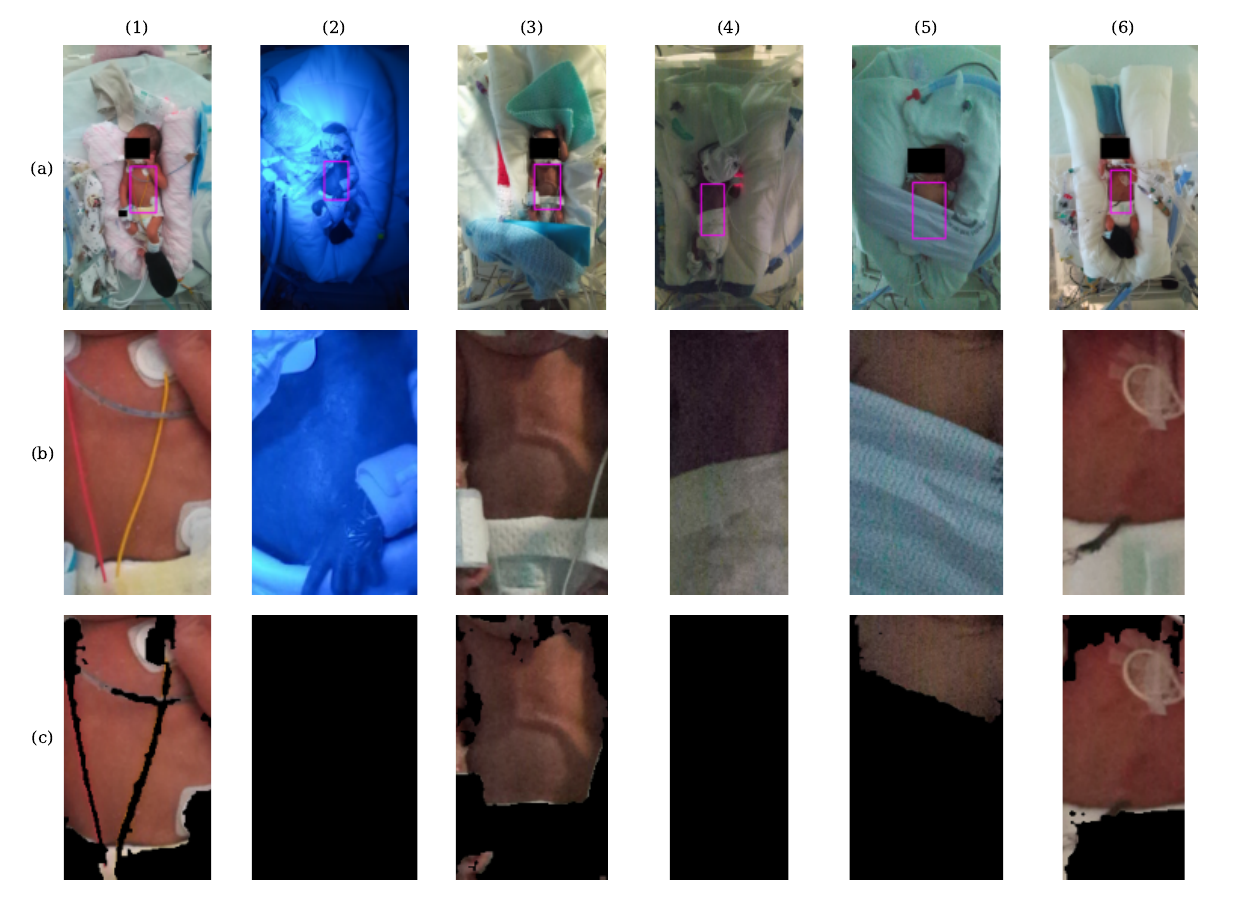}
    \caption{\textbf{Region of interest identification in images.} (a) Representative examples of hip and shoulder identification using the pose estimation algorithms for $6$ samples. (b) Resulting ROIs using hip and shoulder key-points. (c) Subsequent skin segmentation for colour signals. Blue light phototherapy (2) and very low light conditions (4) result in no skin being identified by the segmentation. }
    \label{fig:skinsegmentation}
\end{figure}

 The collected valid vital sign data is depicted in Figure \ref{fig:groundtruthdata}, demonstrating consistency with anticipated neonatal physiological parameters. All video sequences were also manually examined to exclude segments where the neonates were covered during clinical interventions. Regions of interest were then identified using a top-down pose estimation to locate the position of hips and shoulders~\cite{GraftonWarnecke}. For colour signals, non-skin pixels were excluded using a skin mask~\cite{Djamila2020}. Representative examples showing the performance of the methods are illustrated in Figure  \ref{fig:skinsegmentation}. The skin segmentation was found to work in ambient light conditions only. However, the heart rate and oxygen saturation algorithms assume ambient lighting and were found to produce inaccurate readings in very low light and phototherapy conditions when no skin segmentation was used. Therefore, skin segmentation may serve as a safety feature, alerting to conditions where measurement reliability may be compromised.  Notably, respiratory monitoring remains robust under low light, facilitated by infrared illumination from the camera for time-of-flight recordings, with pose detection maintaining acceptable performance even in challenging lighting environments.\\

\subsection{Respiratory rate and volume}
\begin{figure}
    \centering
    \includegraphics[width=\textwidth]{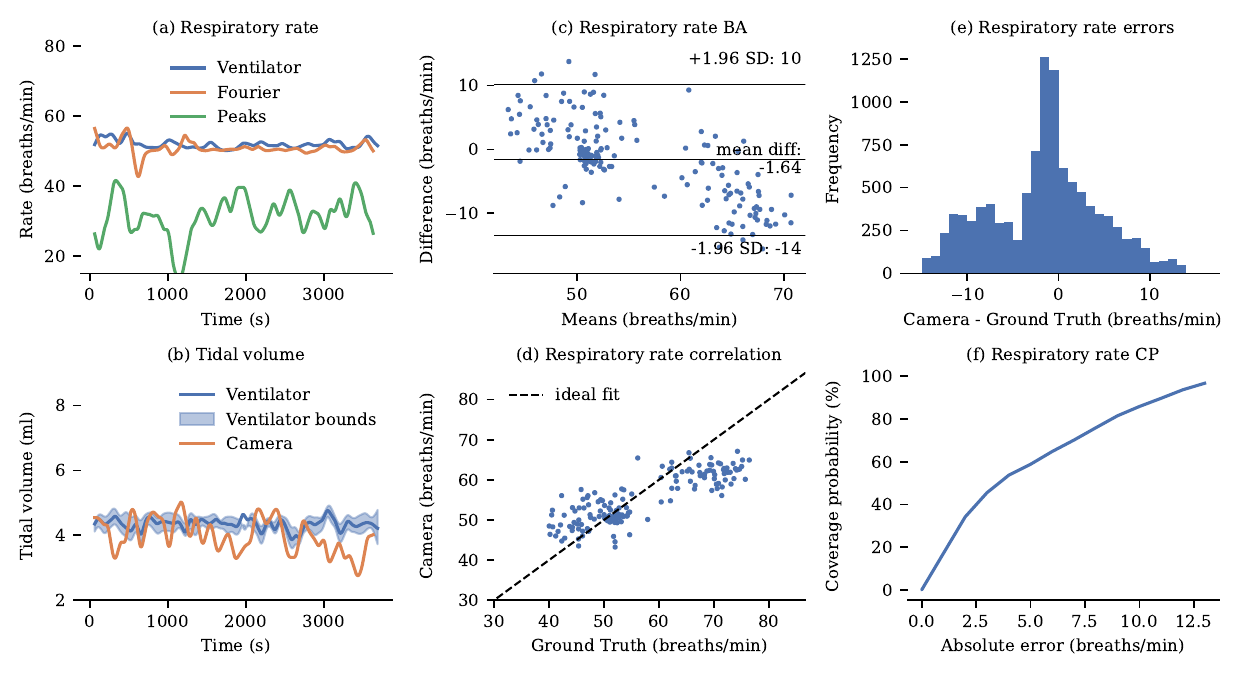}
    \caption{\textbf{Respiratory rate analysis.} (a) Representative $\SI{1}{\h}$ recording of respiratory rate. (b) Representative $\SI{1}{\h}$ recording of tidal volume, illustrating the ventilator ground truth for tidal volumes with the best estimate and associated upper and lower bounds derived from inspiratory and expiratory volumes. Temporary deviations from the ground truth, likely due to infant hand movements, are observed. (c) Bland-Altman plot showing respiratory rate errors. (d) Correlation between camera-derived and ventilator-derived respiratory rates. For clarity, only one sample per minute is displayed in the correlation and Bland-Altman plots. (e) Histogram of error distribution for the camera measurements compared to the ground truth. (f) Coverage Probability (CP) plot showing the proportion of camera-derived values lying within specific absolute error thresholds relative to the ground truth. }
    \label{fig:respiratoryrate}
\end{figure}

 To obtain accurate ground truth data for both respiratory rate and tidal volume, mechanically ventilated babies were included in our study.  The collected RGB-D videos were processed to extract frequency and volume measurements. It is shown that estimating frequency using a Fourier transform of the signal was more accurate than counting the number of peaks (see Table \ref{tab:resp_fourier_peak_counting}).  On the dataset of $3$ ventilated babies, this method achieves an MAE of $4.83$ breaths per minute. A representative example of tidal volume and respiratory rate monitoring is shown in Figure \ref{fig:respiratoryrate}, along with a detailed statistical analysis of the performance of the Fourier transform based respiratory rate estimation in comparison to ventilator ground truth. \\

\begin{figure}
    \centering
    \includegraphics[width=\textwidth]{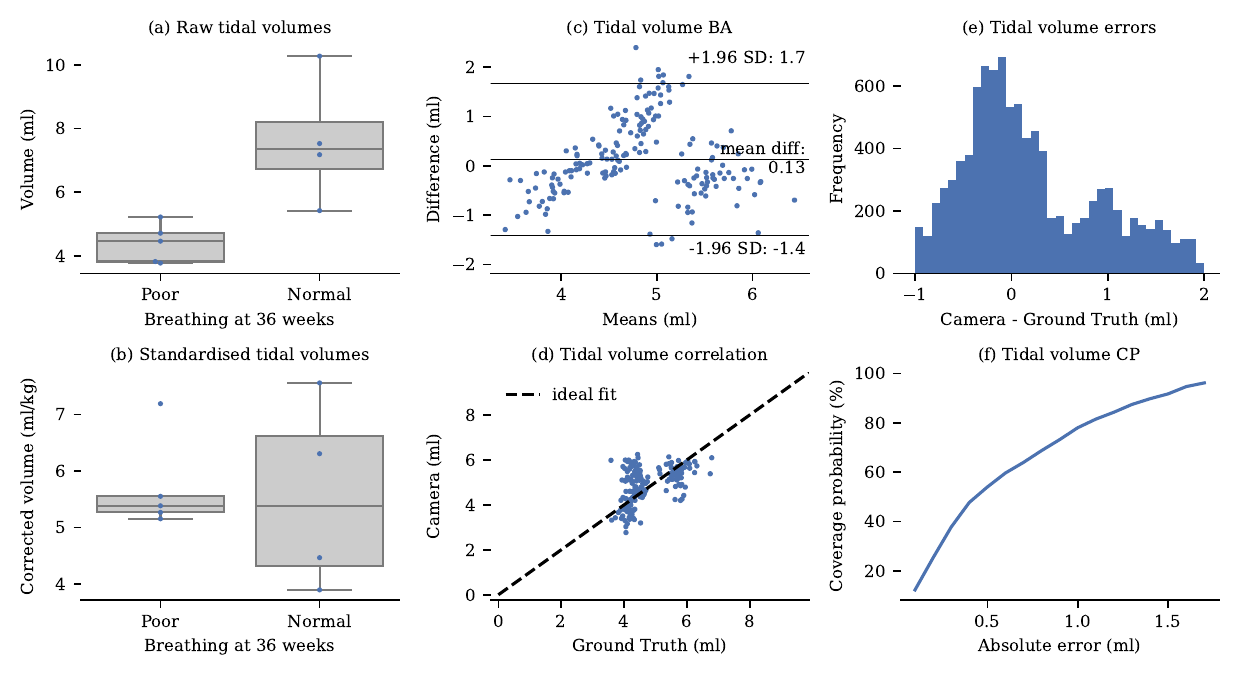}
    \caption{\textbf{Tidal volume analysis.} (a) The normal breathing group has significantly higher $(p=0.016)$ raw tidal volumes than the poor outcome group. (b) Corrected tidal volumes are not significantly different $(p=0.905)$.  (c) Bland-Altman plot showing tidal volume errors. (d) Correlation between camera-derived and ventilator-derived tidal volumes. For clarity, only one sample per minute is displayed in the correlation and Bland-Altman plots. (e) Histogram of error distribution for the camera measurements compared to the ground truth. (f) Coverage Probability (CP) plot showing the proportion of camera-derived values lying within specific absolute error thresholds relative to the ground truth.}
    \label{fig:tidalvolumes}
\end{figure}

Tidal volumes were derived from the depth signal by analysing peak-to-valley differences. The MAE compared to the ventilator's best estimate was found to be $\SI{0.85}{\ml}$ $(17.46 \%)$. Considering potential inaccuracies in the ventilator measurements to define upper and lower bounds on the ground truth based on inspiratory and expiratory volumes, the MAE decreased to  $\SI{0.61}{\ml}$ $(12.81 \%)$ (see Table \ref{tab:tv_table}). The tidal volume measurement accuracy is analysed in Figure \ref{fig:tidalvolumes}.\\

Due to limited availability of ventilated babies in the NICU, our algorithms were also tested on neonates not receiving respiratory support. For these babies, established ground truth values were not available. Instead, the $9$ neonates recorded for approximately $\SI{1}{h}$ each were divided into two groups based on their clinical outcomes after their NICU stay. The poor outcome group comprised babies who either required supplementary oxygen at home or unfortunately did not survive until 36 weeks. Conversely, all babies in the normal breathing group survived and did not require supplementary oxygen after 36 weeks. Babies who exhibited normal breathing at 36 weeks were found to have significantly higher tidal volumes $(p=0.016)$ than their counterparts using a Mann-Whitney U-test (see Figure \ref{fig:tidalvolumes}). However, the tidal volume per kg of body-weight standardisation for neonates, did not differ significantly between groups $(p=0.905)$. This suggests that the algorithms are capable of identifying the higher tidal volumes in the normal group.\\

\subsection{Respiratory dynamics}
\begin{figure}
    \centering
    \includegraphics[width=\textwidth]{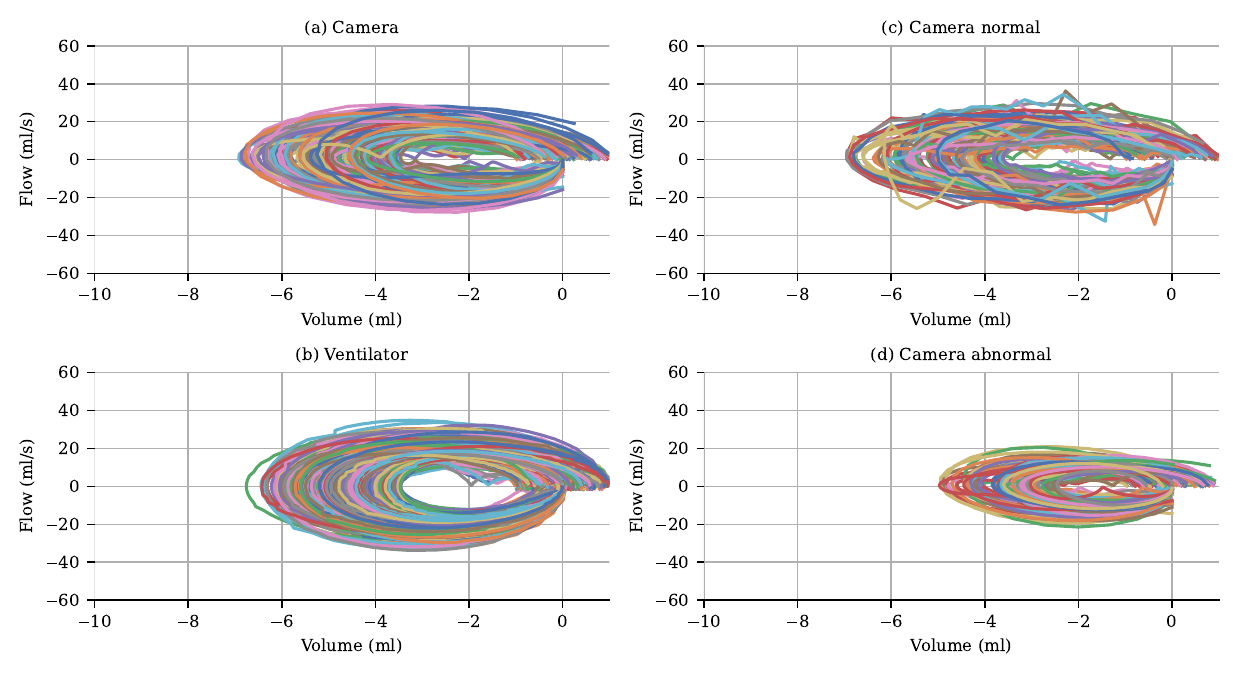}
    \caption{\textbf{Flow-volume loops.} (a) Flow-volume loops for a representative neonate generated using the RGB-D camera, illustrating the flow and volume dynamics during individual breaths. (b) Corresponding flow-volume loops generated from ventilator-derived data for the same neonate. Each loop represents a single breath, showcasing the close correspondence between camera-based and ventilator-based measurements. (c) Flow-volume loops from a non-ventilated neonate in the normal breathing group, demonstrating typical respiratory patterns. (d) Flow-volume loops from a non-ventilated neonate in the poor breathing group are visible smaller.}
    \label{fig:fvloops}
\end{figure}

Flow-volume loops were successfully constructed both from camera and ventilator data. A representative example is shown in Figure \ref{fig:fvloops}. The loops from the two modalities show similar characteristics. Small differences in shape are likely attributed to air leakage in the ventilator-constructed loop. As seen in Figure \ref{fig:respiratoryrate}, the magnitude of this leakage is approximately $\SI{0.5}{\ml}$. During examination of single breaths, it was noted that the camera system was sensitive enough to capture subtle variations in breathing such as interrupted breaths. Interruptions occur when a neonate attempts to breathe during a ventilator-induced breath, which is then terminated prematurely. Examples of a representative normal breath and a prematurely terminated breath followed by a second autonomous breath are shown in Supplementary Information Figure 1 \& 2.  The flow-volume construction was also trialled on non-ventilated neonates, revealing more variable loops characterised by a less smooth pattern (see Figure \ref{fig:fvloops}). In comparing the loops obtained in non-ventilated neonates of the normal breathing group with those of the poor outcome group, the former exhibited larger loops, consistent with the observed differences in tidal volumes. The presented loop from the normal breathing group also has a shape resembling convex expiratory loops documented in the literature. This loop form has been observed in about 1 in 4 neonates that are healthy or suffer from chronic lung disease~\cite{Schmalisch2005}. \\

Using the spatial information gained from the imaging system, it was also possible to examine regional variations in breathing. For this the tidal volumes were calculated using the spatial average of the defined region only. The sensitivity of this approach was demonstrated by showing that the measured regional tidal volumes are higher in uncovered regions than regions partially occluded by medical equipment (see Supplementary Information Figure 3). \\

\subsection{Heart rate}

\begin{figure}
    \centering
    \includegraphics[width=\textwidth]{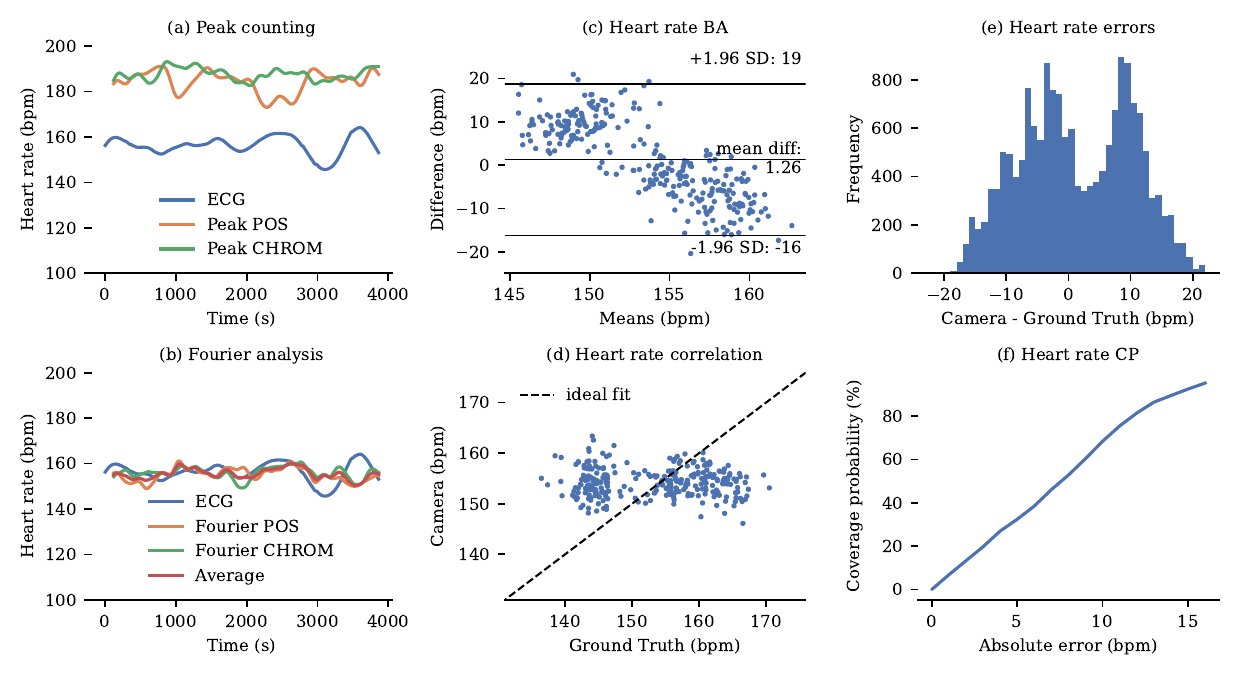}
    \caption{\textbf{Heart rate analysis.} (a) Representative $\SI{1}{\h}$ recording of heart rate monitoring using CHROM and POS signals with peak counting.  (b) Representative $\SI{1}{\h}$ recording of heart rate monitoring using Fourier analysis of the CHROM and POS signals. (c) Bland-Altman plot showing heart rate errors. (d) Correlation between camera-derived and pulse oximeter-derived tidal volumes. For clarity, only one sample per minute is displayed in the correlation and Bland-Altman plots. (e) Histogram of error distribution for the camera measurements compared to the ground truth. (f) Coverage Probability (CP) plot showing the proportion of camera-derived values lying within specific absolute error thresholds relative to the ground truth.}
    \label{fig:heartrate}
\end{figure}

 Five different methods for estimating heart rate from generated colour signals using the CHROM and POS algorithms were trialled (see Table \ref{tab:heart_rate_table}). The signals generated using each algorithm were analysed for frequency using both peak counting and Fourier analysis of the frequency spectrum. Fourier analysis was found to be more accurate than peak counting. Fourier-based methods achieved good accuracy for both the POS and CHROM algorithms, but the best performance was achieved by averaging the heart rate estimations of the two methods. A representative $\SI{1}{\h}$ recording and estimates from all methods are presented in Figure \ref{fig:heartrate}. Across our dataset of $11$ neonates the average of the CHROM and POS algorithms with Fourier analysis achieved a MAE of $\SI{7.69}{\bpm}$. A detailed statistical analysis of the heart rate estimates is provided in Figure \ref{fig:heartrate}.\\

\subsection{Oxygen saturation}

\begin{figure}
    \centering
    \includegraphics[width=\textwidth]{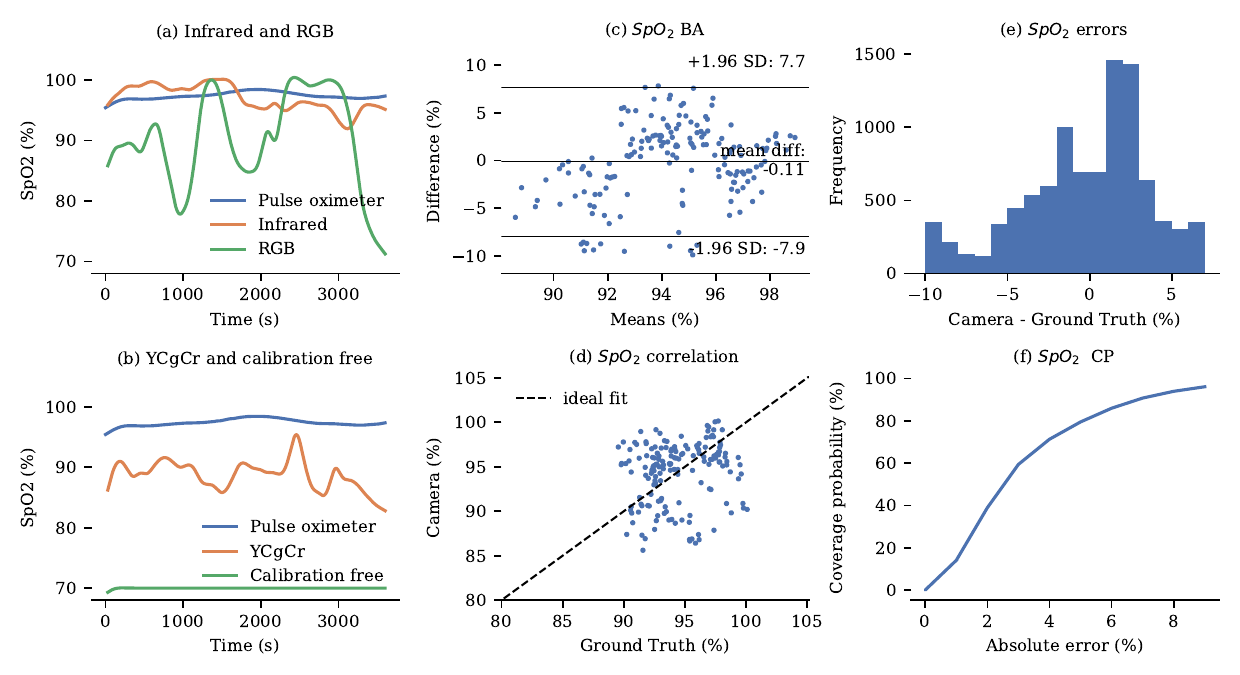}
    \caption{\textbf{Oxygen saturation.} (a) Representative $\SI{1}{\h}$ recording of oxygen saturation monitoring using the infrared and RGB algorithms.  (b) Representative $\SI{1}{\h}$ recording of oxygen saturation monitoring using the  YCgCr and calibration free algorithms. (c) Bland-Altman plot showing oxygen saturation errors. (d) Correlation between camera-derived and pulse oximeter-derived tidal volumes. For clarity, only one sample per minute is displayed in the correlation and Bland-Altman plots. (e) Histogram of error distribution for the camera measurements compared to the ground truth. (f) Coverage Probability (CP) plot showing the proportion of camera-derived values lying within specific absolute error thresholds relative to the ground truth.}
    \label{fig:oxsat}
\end{figure}

For our study we evaluated four different non-contact algorithms that have been reported in the literature. Among these methods, the infrared-based algorithm demonstrated the highest accuracy, as detailed in  Table \ref{tab:ox_sat_tab}. A representative 1h recording comparing all methods is shown in Figure \ref{fig:oxsat}.  Across our dataset of 7 neonates, a MAE  of $3.37\%$ and a MSE of $15.89\%$ were achieved. Notably, the YCgCr colour method also achieved a MAE below $4\%$, and unlike the infrared-based method, it can be achieved using regular RGB cameras. The RGB method performed slightly worse, likely due to the challenging lighting conditions. The calibration-free method was found to introduce significant errors into the measurements. A detailed statistical analysis of the infrared-based method is presented in Figure
\ref{fig:oxsat}.\\

\section{Discussion}
This study introduces the first system  for continuous non-contact estimation of all vital signs considered standard of care.  Importantly, unlike Villaroel et al.~\cite{Villarroel2019}, our method does not require any permanent modification to the incubator. Monitoring was undertaken using a single low-cost RGB-D camera placed above the incubators. The camera is able to record respiratory rate, heart rate and oxygen saturation in line with clinical standards. For the recorded tidal volumes and flow dynamics there are no clinical standards to the best of our knowledge but our results are in good agreement with ground truth. All algorithms were validated in a clinical study in a NICU.\\

Over the past decade, there have been multiple attempts to develop a non-contact RGB-D camera-based respiratory monitoring system for neonates. Cenci et al.~\cite{infantrespmonitoring}  used RGB-D cameras to monitor respiratory rates of NICU patients and compared these values against reference standards. Their study involved $3$ infants, each of which was recorded for a total of $\SI{150}{\s}$. Respiratory rate was extracted from manually defined ROIs by calculating the time differences between adjacent peaks. Although a detailed performance analysis was not provided, the study reports a high correlation  between the respiratory rates obtained from the camera and those derived from a cardiomonitor~\cite{infantrespmonitoring}.\\

Kyrollos et al.~\cite{Kyrollos2021} conducted a more sophisticated study involving the application of deep learning techniques for respiratory monitoring using RGB-D camera data. Their approach entailed training a deep learning network to accurately identify facial and chest regions within the images. Subsequently, respiratory rate was estimated based on signal changes in both RGB and depth data obtained from these identified regions. The algorithm was only validated on a single patient, for which data was recorded for $\SI{20}{\min}$. Although a comprehensive performance analysis was not provided, the study reports that the measurement error for respiratory rate estimation was below $3.5$ breaths per minute for approximately $69\%$ and $67\%$ of the recording duration, when derived from RGB and depth image data respectively~\cite{Kyrollos2021}. \\

Villarroel et al.~\cite{Villarroel2019} conducted an extensive study involving $30$ subjects with recordings spanning over $\SI{400}{\h}$ to monitor respiratory rate. They trained a deep neural network to identify whether babies are present in the image and segment visible skin regions. However, due to the inaccuracies of thoracic impedance measurements only $44\%$ of the recording was included in the analysis. The authors performed a more detailed analysis of their proposed algorithms which achieved a MAE of $3.5$ breaths/min for $82\%$ of the recording time  they considered their ground truth and camera data valid.  Khanam et. al~\cite{Khanam2021} also used deep learning techniques to estimate respiratory rate from $\SI{10}{\s}$ recordings of neonates using two RGB cameras. After exclusion of invalid segments, they achieved a MAE of $2.13$ breaths per minute. \\

Defining accurate requirements for respiratory rate and tidal volume measurements in neonatal monitoring presents challenges due to the absence of consensus on gold standard measurement devices. Currently, there is no universally agreed upon standard for assessing tidal volume measurements.  However, a recent report by UNICEF indicated that achieving a MAE of $5$ breaths per minute would be desirable for low cost infant monitoring~\cite{UNICEF}. The true performance of previously developed monitoring systems remains unclear due to several factors. 
Many studies have provided only superficial analyses, lacking comprehensive validation against established benchmarks~\cite{Kyrollos2021},~\cite{infantrespmonitoring}. Many studies also only recorded under non-natural conditions~~\cite{Rehouma2017}~\cite{Benetazzo2014}, with modifications to the clinical environment~\cite{Villarroel2019} or for very short time segments~\cite{Khanam2021}. Additionally, some studies only included segments of data where their algorithms perform well in the analysis~\cite{Villarroel2019},~\cite{Kyrollos2021}. \\

The accuracy of previous studies in monitoring respiratory parameters is also severely limited by the accuracy of gold standard signals from thoracic impedance measurements~\cite{Villarroel2019}. A more accurate respiratory signal can be obtained from mechanically ventilated neonates, where breath timings are precisely controlled by the ventilator. Usually, when the infants are mechanically ventilated, the impedance recordings are not performed. For one neonate in our study, both impedance and ventilator data were available, suggesting a MAE of 18.5 breaths per minute. MAEs between thoracic impedance and breathing rates counted manually by clinical staff of over $10$ breaths per minute have been reported in the literature~\cite{Jorge2019}. Consultant neonatologists in the Rosie hospital in Cambridge have also expressed reservations about using impedance-derived respiratory measurements to guide clinical decisions due to concerns about accuracy.   \\

We demonstrate that reliable respiratory rates estimates can be obtained from depth signals. On the limited data available, the RGB-D camera outperforms reported accuracies of currently employed thoracic impedance systems. These results align well with previously reported accuracies in the field. Our system has a slightly lower accuracy over the entire dataset than some of the other studies.  This discrepancy is anticipated as we did not exclude large segments of data where the system exhibited lower accuracy, opting instead to provide a comprehensive assessment of performance. Unlike thoracic impedance systems, our approach is also capable of measuring  tidal volumes which can highlight breathing abnormalities. \\

The studies discussed thus far have primarily focused on respiratory rate as a key parameter for assessing respiratory function. However, other parameters such as tidal volumes and flow dynamics are important as well. Tidal volumes in ventilated babies over $\SI{4.5}{\ml}$, for instance, have been shown to be a predictor of successful extubation~\cite{Dassios}. A study by Rehouma et al. measured both respiratory rate and tidal volumes in a paediatric ICU setting~\cite{Rehouma2017}. However, they only imaged manikins with test lungs connected to ventilators rather than actual infants. In a subsequent study by the same group, actual ventilated infants were included, but only one infant was imaged for a total duration of $\SI{5}{\min}$~\cite{REHOUMA2018}. The baby was $4$ months postpartum and had a tidal volume of around $\SI{40}{\ml}$, significantly larger than the neonates in our study, which they could measure with a mean relative error of $9.17\%$. In our dataset, which included three mechanically ventilated premature infants, we observed a slightly higher error rate of $12.81 \%$. This increase is anticipated, as larger tidal volumes facilitate the separation of breathing movements from measurement noise. Our recording also encompassed periods of neonatal movement, which further impacted measurement accuracy but is expected in real-world applications. Furthermore, we demonstrate the non-invasive construction of flow-volume loops that visually closely resemble those recorded by ventilators, a similarity confirmed by consultant neonatologists.  In non-ventilated babies, significantly higher tidal volumes are recorded in neonates with normal breathing compared to those with poor outcomes. The resulting flow-volume loops in the poor outcome group are also smaller.\\

Additionally, we anticipate that the camera could provide valuable insights into respiratory dynamics by capturing spatial variations in breathing efforts and enabling the construction of flow-volume loops. These insights may offer clinicians valuable information to better understand and manage respiratory dynamics in neonates. Tidal flow-volume loops can reveal differences between healthy infants and those with chronic lung diseases~\cite{Schmalisch2005}. Asymmetry in chest wall movements is an early predictor of pneumothorax in neonates, as demonstrated by Wasiman et al.~\cite{Wasiman2016}. Using a sensor attached to the chest, the authors detected changes in chest movement $\SI{31}{\min}$ before a clinical diagnosis of pneumothorax could be confirmed. Another condition that can present with tidal volume asymmetry is congenital scoliosis, as reported by Redding et al.~\cite{Redding2008}. In the future, there is potential to use RGB-D cameras to verify correct endotracheal tube placement in neonates. Endotracheal tube misplacement can have serious consequences including  atelectasis (lung collapse) of the non-ventilated lung, pneumothorax and even death~\cite{Simons2017}. Currently, x-rays are used if tube misplacement is suspected~\cite{Pinheiro2023}. By monitoring changes in spatial airflow distribution, particularly asymmetries that may indicate inadequate ventilation of one lung, RGB-D camera-based systems could offer a real-time method to assess endotracheal tube position without the need for invasive or radiation-based procedures.\\

An application of RGB cameras for heart rate monitoring in a NICU setting was done by Chen et al.~\cite{Chen2020}.  Heart rate was estimated using Fourier analysis of the green channel. On a clinical study of $5$ babies they achieved a MAE of $\SI{7.4}{\bpm}$. Paul et al.~\cite{Paul2020} also used RGB cameras with manual ROI selection. They reported errors of less than $\SI{3} {\bpm}$ in infants with baseline heart rates of about $\SI{120}{\bpm}$ during high-quality recordings, but this level of accuracy was achievable for less than $40\%$ of the recording time. A study by Svoboda et al. has achieved high accuracy in heart rate measurement compared to a pulse oximeter, but this is not as accurate as an ECG~\cite{Svoboda2024}. The previously discussed study conducted by Villaroel et al. also examined heart rate, achieving a MAE of $\SI{2.3}{\bpm}$ over $76\%$ of valid data. Unlike the thoracic impedance measurement, ECG provides robust measurements, of which the authors deemed $91.2\%$ to be valid \cite{Villarroel2019}.\\

Heart rate estimation from colour signals is shown to be feasible for continuous monitoring using our RGB-D camera. The American National Standards Institute specifies an acceptable range of error for heart rate measurements as $\pm 10\%$  or $\pm \SI{5}{\bpm}$, whichever is greater~\cite{AAMI2002}. However, higher accuracy has been achieved in the literature using deep learning approaches  that are less sensitive to variations in lighting conditions and more expensive cameras. This difference is likely attributed to the high noise inherent in colour signals, in contrast to the more robust depth measurements obtained using built-in infrared lighting. On a similar dataset, Grafton et al. ~\cite{GraftonWarnecke} have achieved an MAE of $\SI{3.83}{\bpm}$ with deep learning methods. A large scale study conducted by Huang et al.~\cite{Huang2021} used deep learning techniques for estimation of heart rate in neonates, achieving a MAE of $\SI{3.97}{\bpm}$. The authors from the latter study also directly compared their method to established heart rate estimates and concluded that it outperformed the CHROM and POS algorithms slightly on their dataset. However, their recording session only lasted a maximum of 1 min, leaving uncertainty about the system's accuracy for long-term monitoring applications. \\

Various established techniques for measuring oxygen saturation using cameras have been documented in the literature~\cite{oxygeninfrared},~\cite{Guazzi2015},~\cite{YCgCR_Kim},~\cite{Sasaki}. In laboratory settings they have been reported to give good accuracy but there is limited evidence on their performance in challenging clinical environments. One of the first applications of non-contact monitoring on neonates was performed by Ye et al. \cite{Ye2024}. They managed to record oxygen saturation from $22$ infants on the NICU with a MAE of less than $4\%$. Villaroel et al.~\cite{Villaroel2014} previously demonstrated the capture of a bradycardia episode resulting in desaturation using an RGB camera. They, however, did not provide any statistical analysis of their method in the study.\\

For our study, we evaluated four different non-contact algorithms that have been reported in the literature. The superior performance of the infrared based method is not surprising given that traditional pulse oximeters have the same principle of operation. The infrared result is compliant with UK regulations that mandate the MSE to be below $16\%$~\cite{UKGovPulseOx}. The RGB method performed slightly worse, likely due to the challenging lighting conditions. The calibration-free method was found to introduce significant errors into the measurements. This is likely due to the inadequacy of the shallow and deep layer model for neonates or the inaccurate modelling of our camera  under the assumption of absorbance at a single wavelength. The model is also based on the absorption characteristics of adult haemoglobin Hb A. Neonates, however, mainly have Hb F, which was produced during the pregnancy and remains the dominant form until $6-12$ months post delivery~\cite{Kaufmann2022}. Despite structural disparities, studies have indicated negligible differences in oxygen absorption values between fetal and adult haemoglobin molecules~\cite{Harris1988}. A limitation of our study is that we only compared the $SpO_2$ to devices used in clinical practice. For full validation,  establishing ground truth using blood gas samples would be necessary. However, this procedure is highly invasive, only provides intermitted measurements, and falls beyond the scope of our study.\\

In contrast to previous studies, performance of the developed algorithms was evaluated on the entire recorded dataset where the baby is visible in the image rgardless of the baby's position and valid ground truth data was available. Additionally, we have created a fully anonymised dataset of the pre-processed data that can be made publicly available along with the code used in the analysis. This may facilitate the development of novel non-contact monitoring algorithms, addressing the current limitation of publicly available neonatal data for testing such methods.\\

 Future efforts will focus on validating the algorithms on a larger study population. Specifically, inclusion of neonates with respiratory pathologies will be crucial to verify the diagnostic potential of the system. By providing additional clinical insights, currently not available to clinicians, the system could assist in early recognition of pathological changes, with the potential to improve patient outcomes. The non-contact vital sign monitoring system may integrate into a broader camera system for neonatal activity and physiology monitoring currently under development. \\

\section{Methods}
\subsection{Clinical study}
  Ethical approval was obtained for a single-centre study conducted at Cambridge University Hospitals NHS Foundation Trust Rosie Hospital, involving the monitoring of neonates using RGB-D cameras to assess physical activity and vital functions. The study was reviewed by the Research Ethics Committee (North-West, Preston) with reference number 21/NW/0194 and IRAS ID 285615. Ground truth measurements were concurrently recorded using gold standard devices in the ICU including a pulse oximeter, ventilator, ECG, and thoracic impedance. All collected data underwent anonymisation following the extraction of relevant clinical information from hospital records to protect patient privacy.  Informed written consent was obtained from all parents for the recording and subsequent analysis of the data. Participants had the right to withdraw from the study at any time and request deletion of their collected data.\\

The study was designed to minimise interference with clinical interventions and parental interactions with the babies.  The recording setup was non-invasive and did not require any permanent changes to the incubator, ensuring it could be quickly removed during emergencies. Throughout the study, the clothing and positioning of the infants remained unchanged. For respiratory monitoring we specifically targeted babies on mechanical ventilation. During data collection, ventilated babies were recorded for up to $\SI{1}{\h}$, while non-ventilated babies were monitored for up to $\SI{72}{\h}$ for physical activity monitoring. For this analysis, only infants who were nursed naked were included as visible skin on the chest area is required for the developed algorithms. Neonates receiving mechanical ventilation were only included in the study if it was clinically appropriate to record them in a supine position.\\
\subsection{Experimental setup}
\subsubsection{RGB-D camera}
The RGB-D camera used in this study was the  Azure Kinect DK (Microsoft Corporation, USA), equipped with a 1MP Time-of-flight (ToF) depth camera and a 12MP CMOS RGB camera~\cite{AzureKinect}. Colour images with a resolution of $1280\times720$ were captured with $4:2:2$ chroma sub-sampling and stored in the MJPEG format. The depth image acquisition relied on the Amplitude Modulated Continuous Wave (AMCW) time-of-flight principle which involves illuminating the scene with near-infrared light emitted by laser diodes~\cite{Kinectdocs}. The depth image is calculated by measuring the time difference between the start of illumination and the return of the infrared light from specific objects. Additionally, an infrared image is obtained by measuring the intensity of the reflected light. The infrared and depth images were upsampled from $640\times576$ to match the dimensions of the RGB image. The Azure Kinect offers both narrow and wide field of view options, with the narrow field of view mode selected for this study. \\

\subsubsection{Data collection in the NICU}
The camera was placed directly on top of the incubator's perspex lid (see Fig. \ref{fig:setup}). To securely mount the camera, a Manfrotto Flexible Arm was used, which was clamped onto the incubator. The arm allowed for easy adjustment and movement of the camera, facilitating quick repositioning
during clinical emergencies.   Heart rate was measured using a single-lead ECG signal, sampled at $\SI{240}{\Hz}$, while oxygen saturation data were obtained from standard clip-on pulse oximeters used on the ward. The patient monitor (GE HealthCare Technologies Inc., USA) collecting heart rate and oxygen saturation  was connected to a computer via an Ethernet cable and data were recorded using a custom software developed in the Signal Processing and Communications lab at the Department of Engineering, University of Cambridge. Respiratory data were collected by linking a laptop directly to the ventilator, which recorded ventilation parameters using software provided by the manufacturer (Dräger Medical GmbH). During data collection, it was noted that the introduction of the infrared light from the camera affected the oxygen saturation probe that relies on infrared light to estimate the bound haemoglobin~\cite{Dosso2021},~\cite{Grafton}. To mitigate this interference, the pulse oximeter was shielded with a phototherapy mask, designed to block external light sources (see Fig. \ref{fig:setup}) and the SpO2 signal manually observed before and after activating the infrared emitter. This precautionary measure ensured reliable oxygen saturation measurements despite the presence of infrared illumination. \\

\subsection{Data analysis}\label{sec:data_analysis}
\begin{figure}
    \centering
    \includegraphics[width=\linewidth]{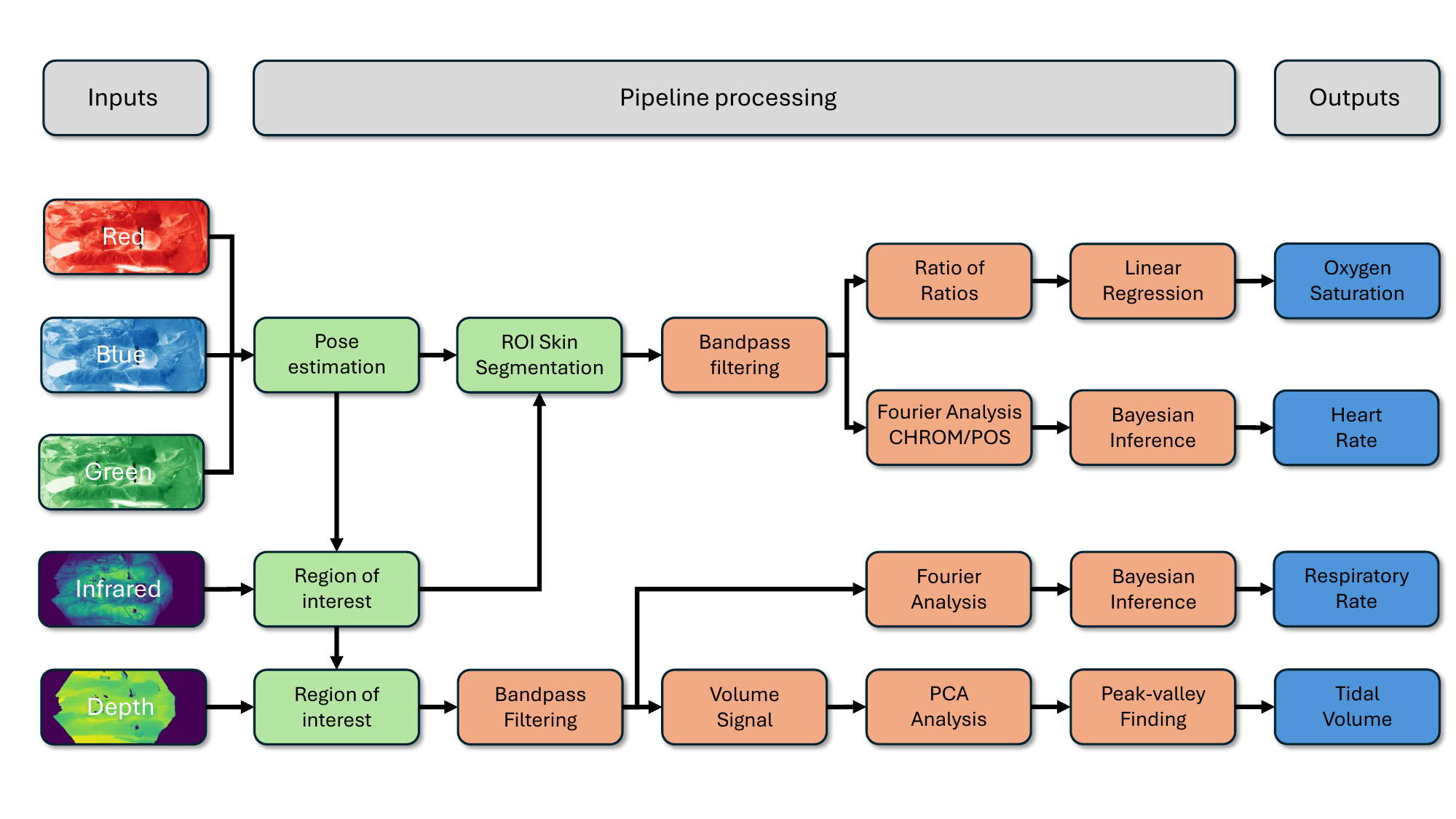}
    \caption{\textbf{Pipeline for determining vital signs from RGB-D videos.} The red, green, and blue channels are utilised for pose estimation and skin segmentation, defining the regions of interest. The resulting average signals from these regions undergo Bandpass filtering. Oxygen saturation is calculated using a linear regression of the ratio of ratios. Heart and respiratory rates are estimated via Bayesian inference applied to their respective Fourier spectra. Volumes are estimated from peak-to-valley measurements in the volume signal. Processing operations involving a single frame are indicated in green, while those involving time-series signals are shown in light red. }
    \label{fig:pipeline}
\end{figure}

All data analysis was conducted using Python (version 3.11.5). Camera and patient monitor data were analysed using
NumPy (version 1.25.2), SciPy (version 1.11.2), scikit-learn (version 1.3.1), and simdkalman (1.0.4). Analysis of ventilator data were conducted using ventiliser (version 1.0.0). Matplotlib (version 3.8.0), and seaborn (version 0.13.0) were used for visualisation. An overview of the data pipeline is shown in Figure \ref{fig:pipeline}.

\subsubsection{Region of interest identification}
The analysis focused on the thoracic and abdominal regions, defined by the hip and shoulder joints, a method successfully employed in the literature~\cite{Benetazzo2014}. Neonatal pose estimation relied on the MMPose toolbox~\cite{MMPose}, where a standard model was retrained specifically for neonatal applications using a manually annotated dataset featuring keypoints for hips and shoulders~\cite{GraftonWarnecke}. For the colour-based heart rate and oxygen saturation measurements, non-skin pixels were excluded to enhance signal quality. Skin segmentation was achieved by isolating pixels in the YCbCr and HSV colour spaces~\cite{Djamila2020}. The skin masks generated from different color spaces were processed to improve their accuracy and consistency. First, each mask underwent morphological opening using a 3x3 kernel to reduce noise by eliminating small, isolated regions or filling tiny gaps. Next, the processed masks from all color spaces were intersected. The resulting combined mask was then smoothed using a median filter of size 3 to reduce remaining noise. Finally, the mask was refined with another morphological opening operation, this time using a larger 4x4 kernel.  \\
\subsubsection{Signal extraction}
To extract a one-dimensional signal from the data for subsequent processing, the pixels within the region of interest were spatially averaged, with zero values (indicating invalid measurements) excluded from the averaging process. The colour signals obtained were sufficiently clear for further analysis, as the skin mask effectively removed noisy portions of the torso region. However, for the respiratory signal, using the skin mask did not produce optimal results because areas covered, for instance by electrodes, constitute important parts of the signal. This approach renders the respiratory signal susceptible to artefacts stemming from the variable inclusion of edge regions, especially under conditions of movement. \\

To enhance the accuracy of signal extraction, edge and outlier regions were excluded from the averaging process by analysing the depth values distribution in a histogram. Pixels deviating more than $\SI{25}{\mm}$ from the values of the most frequent bin, corresponding to the main torso region, were disregarded.  Deviations in neonatal chest depth exceeding $\SI{25}{\mm}$ are highly improbable and likely represent pixels from blankets, medical equipment, or limbs rather than the torso itself. This results in a clearer signal (see Fig. \ref{fig:sigproc}). \\
\subsubsection{Respiratory monitoring}
\begin{figure}
    \centering
    \includegraphics[width=\textwidth]{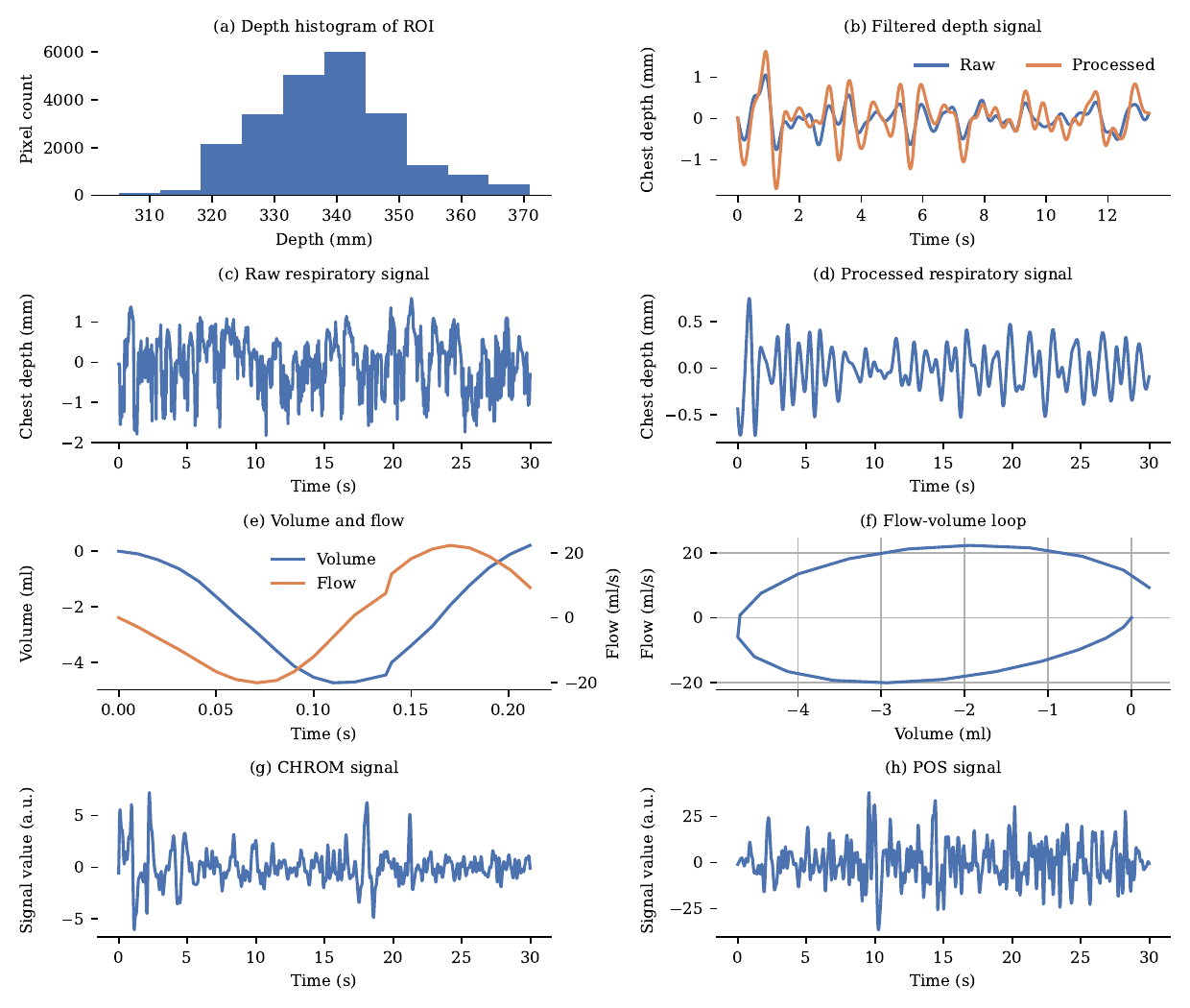}
    \caption{\textbf{Signal processing.} (a) Representative depth histogram of a region of interest for a single frame. (b) Outlier pixels deviating by more than $\SI{25}{\mm}$ from this peak are excluded from further processing to reduce noise. (c) Representative $\SI{3}{\s}$ raw respiratory signal. (d)  Bandpass filtering and PCA remove high frequency noise effectively and result in a signal with visible peak. (e) Construction of a flow-volume loop from camera data involves differentiating the captured volume signal to obtain a flow signal, which represents the rate of change of volume over time. (f) Flow and volume signals are combined in a flow-volume loop to visualise their relationship during a single respiratory cycle. (g) Representative $\SI{3}{\s}$ CHROM signal. (h) Representative $\SI{3}{\s}$ POS signal. The CHROM and POS signals are a combination of the dimensionless colour intensity signals and are displayed in arbitrary units (a.u.).}
    \label{fig:sigproc}
\end{figure}

The resulting raw signal was then processed using a $7^{th}$ order Butterworth bandpass filter and a principal component  analysis (PCA), as illustrated in Figure \ref{fig:sigproc}. The cuoffs for the Butterworth filter were chosen at $15$ and $150$ breaths per minute which includes the entire physiological range for neonates~\cite{Fleming2011}. All filtering procedures in the study were performed bidirectionally (both forwards and backwards) to mitigate any potential shifts in the signal. For real-time analysis, only forward filtering would be applied and the associated shift accounted for. The PCA was performed by converting the one-dimensional time series signal into a Hankel matrix $H$ of dimension $k\times l$ 
\begin{equation}
    H=\begin{pmatrix} t_1 & t_2 & \dots &t_k\\
    t_2 & t_3 &... & t_{k+1}\\
    \vdots & \vdots & \ddots & \vdots\\
    t_l & t_{l+1} & ... & t_{n}
    \end{pmatrix},
\end{equation}
where $t_i$ are elements of a time series $T=(t_1, t_2 \dots t_n)$ \cite{RODRIGUES2013}. To include all values in the matrix we set $k=n-l+1$. Subsequently, a singular value decomposition (SVD) of the matrix is performed. Optimal results were achieved by retaining the first principal component of the volume signal and the first five components of the respiratory rate signal as some of the higher frequencies were not captured by the first principal component. After recalculating the matrix, the time series can be recovered by averaging the columns, accounting for their offsets. It was noted that this approach was still susceptible to parts of the thorax and abdomen being covered for instance by clinical equipment or hand movement.\\ 

To address the challenge of obscured regions in the signal, the region of interest was divided into four quadrants. Each quadrant was spatially averaged, and the resulting signals were processed using the described methodology involving Butterworth filtering and PCA. For each time point it was then determined which of the quadrants had a valid signal by considering the physiological range of neonatal tidal volumes. The valid depth values for each time point were then averaged.\\

The resulting respiratory signal can be used to estimate respiratory rate using two approaches: Peak counting and Fourier analysis of the frequency spectrum. In peak counting, breaths are identified for each $\SI{60}{\s}$ segment (with a $\SI{1}{\s}$ stride) by detecting peaks corresponding to tidal volumes within two standard deviations of the median tidal volume for each infant. This method helps filter out potential movement artefacts by excluding breaths that significantly deviate from typical respiratory patterns. Breaths exceeding $\SI{7.5}{\ml}$ or $1.5$ times the median tidal volume of the signal, whichever is greater, and those smaller than $\SI{2}{\ml}$ are excluded from the standard deviation calculation to ensure accuracy. The dual threshold ensures that all breaths in the expected range of neonates are captured without excluding breaths from larger babies as outliers. Alternatively, examining the frequency spectrum by applying a Fourier transform to the same segments as used for the peak counting can estimate respiratory rate. To refine this analysis, a Hamming window was applied to the signal segments for pulse shaping.\\

Bayesian inference was employed with the Fourier frequency spectrum as the likelihood and a Gaussian prior characterised by a mean of $50$ breaths per minute and a standard deviation of $15$ breaths per minute to enhance the precision of the frequency spectrum estimate. This prior reflects the expected distribution of respiratory rates observed in neonates~\cite{Fleming2011}.  To account for variations in mean breathing rates across individuals, an adaptive strategy was employed. Specifically, the prior distribution was updated by multiplying the original Gaussian prior with a Gaussian estimate derived from at least ten measurements of breathing rate obtained without Bayesian inference, resulting in an unnormalised Gaussian. This iterative approach adjusts the prior based on previously observed respiratory rate data for each individual neonate as well as accounting for the expected value in the population, improving the accuracy of the Bayesian inference process. Both respiratory rate signals were subsequently passed through a Kalman filter with a process noise of 
\begin{equation}
Q=\begin{bmatrix}
10^{-4} & 0 \\
0 & 10^{-5}
\end{bmatrix}
\end{equation}
and an observation noise $R$ of $20$ breaths per minute. \\

To derive a signal for tidal volume measurement, the changes in depth were multiplied by the area of the ROI rectangle. The pixel coordinates $(x,y)$ were converted to world coordinates $(X,Y)$ using the equations
\begin{align}
    X&= \frac{Z(x-p_x)}{f_x},\\
    Y&= \frac{Z(y-p_y)}{f_y},
\end{align}
where $Z$ represents the depth value of the image pixel and $p_x$, $p_y$, $f_x$, $f_y$ are camera  calibration parameters defining the imaging process between world and image coordinates. The resulting X and Y coordinate time series were bandpass filtered with the same filter as the depth signal to reduce noise. The volume signal $V$ is then obtained as
\begin{equation}
    V=Z(X_2-X_1)(Y_2-Y_1),
\end{equation}
where $X_1$, $X_2$, $Y_1$ and $Y_2$ are the upper and lower bounds of the region of interest. The signal was then processed using the bandpass filter and PCA, as for the respiratory rate. The tidal volume was calculated as the mean of the peak-valley difference in a $\SI{60}{\s}$ interval of all the valid peaks identified in the signal. The tidal volume signal was then processed using a Kalman filter with a process noise of 
\begin{equation}
Q=\begin{bmatrix}
10^{-4} & 0 \\
0 & 10^{-5}
\end{bmatrix}
\end{equation}
and an observation noise $R$ of  $\SI{2}{\ml}$.\\

 Even though the ventilator provides a very controlled environment, it is difficult to accurately measure tidal volumes due to leakage. Accurate measurements of the air volume of inspiration and a slightly lower volume of expiration are however available.  In this study, both the most accurate estimate from the ventilator based on leakage modelling and the worst-case estimate, encompassing all values between the inspiratory and expiratory volumes, were considered. This approach accounts for uncertainty regarding the location and extent of any leakage in the system. For accurate comparison, the values obtained from the ventilator for rate and tidal volume were also averaged for $\SI{60}{\s}$ intervals and processed using Kalman filters with the same parameters as for the camera data.\\

The camera data can be used to construct flow-volume loops which are normally obtained from ventilator or spirometry measurements.  The ventilator measures airflow, which can be numerically integrated to generate a volume signal. To ensure accurate comparison between the data collected using two different devices, the flow data from the ventilator is bandpass filtered before integration using the same filter as the respiratory signal derived from the camera. Conversely, the camera captures a volume signal, which can be numerically differentiated to derive a flow signal. The beginning and end of each breath in the ventilator data were identified using a rule-based algorithm involving the analysis of flow and pressure data~\cite{Gusztav}.  An example of a volume signal recorded with the camera for a single breath, along with the resulting flow profile and flow-volume loop, is shown in Figure \ref{fig:sigproc}.  Flow-volume loops where the end point deviated by more than $\SI{1}{\ml}$ from the start volume, the tidal volume was less than $\SI{2}{\ml}$, more than one standard deviation away from the median, or the flow values deviated highly from the expected sinusoidal curve are not visualised.\\

\subsubsection{Heart rate monitoring}
The resulting colour signals obtained from spatial averaging can be combined into a single physiological signal using the CHROM~\cite{CHROM} and POS~\cite{Wang2016} algorithms. These algorithms use empirically determined equations that were shown to yield a high signal-to-noise ratio. For the CHROM algorithm, the mean of each colour channel $C \in (R, G, B)$ is calculated and the channel normalised
\begin{equation}
    C_{norm_i}= \frac{C_i}{\mu(C_i)},
\end{equation}
where $\mu$ represents the  mean of the respective time series. Additionally, better results are achieved by standardising the colour channels 
\begin{equation}\label{eq:CHROM standard}
    [R_{std}, G_{std}, B_{std}]= [0.7682\: R_{norm}, 0.5121\:  G_{norm}, 0.3841\:  B_{norm}],
\end{equation}
with empirically determined coefficients~\cite{CHROM}. The light reflected from the skin is assumed to consist of both broadly scattered diffusion and a narrow specular component.  The specular component only depends on the light source whereas the diffusion component varies with the blood volume. The specular component can be eliminated by constructing orthogonal channels
\begin{align}
    X_{std} &=\frac{R_{std} -G_{std}}{0.7682-0.5121}= 3R_{norm}-2G_{norm},\\
    Y_{std} &= \frac{R_{std}+G_{std}-2B_{std}}{0.7682+0.5121-0.7682}= 1.5 R_{norm} + 0.5 G_{norm}- B_{norm},
\end{align}
from colour difference as the specular reflection will be similar for all channels. The final signal is then calculated as 
\begin{equation}
    S=3\left(1-\frac{1}{2}\frac{\sigma(X_f)}{\sigma(Y_f)}\right)R_f-2\left(1+\frac{1}{2}\frac{\sigma(X_f)}{\sigma(Y_f)}\right)G_f+\frac{3}{2}\frac{ \sigma(X_f)}{\sigma(Y_f)}B_f,
\end{equation}
where $R_f$, $G_f$, $B_f$, $X_f$ and $Y_f$ are obtained by bandpass filtering the colour channels $R_{std}, G_{std}, B_{std}$, $X_{std}$, and $Y_{std}$  and $\sigma$ is the standard deviation of the signal. \\

The POS algorithm begins by spatially averaging the three colour channels for each frame. Subsequently, these averaged values are temporally normalised within specified windows 
\begin{equation}
    C_{norm_{i_{a\to b}}}=\frac{C_{i_{a\to b}}}{\mu(C_{i_{a\to b}})},
\end{equation}
where $a$ and $b$ indicate the start and end of the window, respectively. Similar to CHROM, a set of orthogonal colour channels is then defined: 
\begin{align}
    X_{norm}&=G_{norm}-B_{norm},\\
    Y_{norm}&=G_{norm}+B_{norm}-2R_{norm}.
\end{align}
Each window signal is then calculated as
\begin{equation}
    S= \left( 1+ \frac{\sigma(X_{norm})}{\sigma(Y_{norm})} \right) G + \left(\frac{\sigma(X_{norm})}{\sigma(Y_{norm})}-1\right) B - 2 \frac{\sigma(X_{norm})}{\sigma(Y_{norm})}R.
\end{equation}
The final signal is obtained as a sum of each of the overlapping window signals and subtracting their respective means~\cite{Wang2016}. This process enhances the identification of peaks within the signal. Furthermore, the signal is bandpass filtered using the same filter utilised in the CHROM algorithm to improve peak detection accuracy.\\

To exclude oscillatory signals generated by respiration and remove high frequency noise,  the CHROM and POS signals were processed with a Butterworth bandpass filter with cutoffs of $\SI{90}{\bpm}$ and $\SI{270}{\bpm}$, following methods similar to those in the literature~\cite{Villarroel2019}. Sample recordings from a representative $\SI{30}{\s}$ video using the CHROM and POS algorithms are shown in Figure \ref{fig:sigproc}. These signals can be utilised to estimate heart rate using similar approaches as for respiratory rate estimation but with a window length of $\SI{120}{\s}$.  Bayesian inference was performed using a Gaussian prior with a mean of $\SI{155}{\bpm}$ and a standard deviation of $\SI{15}{\bpm}$ instead, which approximates the expected distribution of heart rate signals in neonates~\cite{Alonzo2018}. Due to the noise in the raw Fourier spectrum, the adaptive prior method was deemed unreliable. The signal was then smoothed using the same Kalman filter applied to the respiratory signal.  \\

When analysing ECG-derived heart rate measurements, it was observed that a small proportion of values exhibited non-physiological jumps that were not plausible given the neighbouring values.  Two representative examples of such  jumps are shown in Supplementary Information Table 1. As these measurements would skew the comparison of camera derived heart rate and ground truth they were manually removed. The valid ECG values were smoothed with the same Kalman filter to make measurements more comparable.\\

\subsubsection{Oxygen saturation monitoring}
The time series data from each colour channel was processed to extract both the alternating current $(AC)$ and direct current $(DC)$ components. Initially, all signals underwent Butterworth bandpass filtering with cutoffs of $\SI{60}{\bpm}$ and $\SI{300}{\bpm}$. Subsequently, the filtered signal was segmented into $\SI{30}{\s}$ intervals. Within each segment, the $AC$ component was determined as the average amplitude of the peaks observed within the interval. Meanwhile, the $DC$ component was estimated as the mean value of the signal over the same interval. The $AC$ and $DC$ components of the RGB and infrared channels were utilised in the calculation of oxygen saturation using four distinct approaches described in the literature. All of these assume a pulsatile component of optical absorption attributed to increases in arterial blood during the systolic phase of the heartbeat, along with a non-pulsatile component originating from other tissues.\\

Similar to pulse oximeters used in the clinic, oxygen saturation has been be monitored by measuring differences in absorption of red and infrared as the absorption at these wavelengths differs depending on the oxygenation state (see Supplementary Figure 4)~\cite{oxygeninfrared}. 
Even in the absence of an infrared signal, a comparable analysis can be conducted, as there are also differences in absorption between red and blue light~\cite{Guazzi2015}.
A different approach developed by Kim et al.~\cite{YCgCR_Kim} calculates the oxygen saturation colour signals in the YCgCr colour space. It is also possible to measure oxygen saturation without performing a linear regression by modelling the light absorption in the skin~\cite{Sasaki}. Obtained signals were then truncated at $70\%$ and $100\%$, as these represent realistic physiological boundaries. Subsequently, the signal was smoothed with a Kalman filter with a process noise 
\begin{equation}
Q=\begin{bmatrix}
10^{-4} & 0 \\
0 & 10^{-5}
\end{bmatrix}
\end{equation}
and an observation noise $R$ of $10\%$. \\

The oxygen saturation measurements from the gold standard pulse oximeter occasionally displayed values that were likely erroneous, suggesting fatal desaturation of the baby that was not observed during the recording period as the baby remained pink and well perfused. To validate the measurements, the underlying waveform was analysed for periodicity. A less periodic signal, often observed during movements, suggested potential measurement errors. Examples of good and bad waveforms are illustrated in Supplementary Information Figure 5. To improve data quality, all oxygen saturation recordings were segmented into $\SI{30}{\s}$ intervals, and segments with low-quality signals were manually excluded. Reasons for segment exclusion included lack of clear oscillations, prolonged periods of constant values and sharp spikes of abnormally high or low values.  The resulting revised curve is depicted for two representative babies in Supplementary Information Figure 6. This approach facilitated the identification and removal of unreliable data segments, enabling a more accurate assessment of oxygen saturation patterns and minimising the impact of movement-related artefacts on the analysis. For accurate comparison, the values were Kalman filtered with the same parameters as the measurements obtained form the camera.\\

\subsubsection{Statistical analysis}
The most commonly used method to assess
differences between gold standard and new methods in clinical medicine is a Bland-Altman plot~\cite{Zaki}. Bland-Altman plots examine the difference between two measurements as a function of the mean of the two measurements~\cite{BlandAltman}. This permits the detection of systematic trends in the error distribution. The bias in the new measurement is approximated by the mean average error. Limits of agreements are drawn at $\pm 2$ standard deviations, providing a range within which most of the differences between measurements fall. The assumption underlying Bland-Altman analysis is that the noise in the measurements is approximately normally distributed. This assumption can be verified by constructing a histogram of the errors. \\

Apart from Bland-Altman plots, there is no established consensus on the most appropriate metrics for comparison of clinical measurements with a variety of methods being used in the literature~\cite{McLaughlin}. Many articles have also been reported to incorporate inappropriate tests such as correlation coefficients or t-tests~\cite{McLaughlin}. Instead, it is recommended to use coverage probability (CP), which measures the proportion of measurements that fall within $\pm \kappa \%$ of the reference signal~\cite{Lawrence2002}. Commonly used metrics such as mean absolute error (MAE) and mean square error (MSE) are also reported.  The statistical methods used are defined in Table \ref{tab:statisticalmetrics}. To match the measurements obtained from different sensors with the camera data, the higher sampled signals were averaged over periods corresponding to the period of the lower sampled signal. The tidal volume comparisons between different groups were done using a two-sided Mann–Whitney U test. \\

\begin{table}[H]
    \centering
    \begin{tabular}{c c}
    \toprule
        \thead{Description} & \thead{Value}  \\
        \midrule
         \makecell{Total number of patients} & 14\\
         \makecell{Total video length (min)} & \makecell{744} \\
         \makecell{Average recording time per patient (min)} & \makecell{$53\pm13$} \\
         \makecell{Gestational age at birth (weeks)} & \makecell{$27\pm2$}\\
         \makecell{Gestational age at recording (weeks)} & \makecell{$29\pm3$}\\
         \makecell{Weight at birth (grams)} & \makecell{$893\pm322$}\\ 
         \makecell{Weight at time of recording (grams)} & \makecell{$1033\pm348$}\\
         \makecell{Gender} & \makecell{Male: $11$ Female: $3$}\\
         \makecell{Ethnicity}& \makecell{White: $8$ Non-White: $6$}\\
         \bottomrule
    \end{tabular}
    \caption{\textbf{Summary of the clinical study population.} Average values are presented with their corresponding standard deviations.}
    \label{tab:patient_demographics}
\end{table}

 \begin{table}[H]
    \centering
    \begin{tabular}{c c c c c}
    \toprule
        \thead{Method}& \thead{$\downarrow$MAE (/min)}&\thead{$\downarrow$MSE (/min)}&\thead{$\uparrow$CP$\pm10\%$ (\%)}& \thead{$\uparrow$ CP$\pm20\%$ (\%)}\\
    \midrule
        \makecell{Peak counting} & \makecell{$23.14$} & \makecell{$651.30$} & \makecell{$2.83$} &\makecell{$12.64$} \\
        \makecell{Fourier} & \makecell{$4.84$}&\makecell{$39.47$}&\makecell{$62.30$}&\makecell{$90.74$}     \\
        
    \bottomrule
    \end{tabular}
    \caption{\textbf{Comparison of respiratory rate estimation methods.} Total number of subjects: $3$. Arrows indicate whether higher or lower values signify increased accuracy. Fourier analysis results in significantly more accurate respiratory rate estimation. The peak counting algorithm is very sensitive to movements.}
    \label{tab:resp_fourier_peak_counting}
\end{table}

\begin{table}[H]
    \centering
    \begin{tabular}{ c c c c c}
    \toprule
       \thead{Ground truth }& \thead{$\downarrow$MAE (ml)} &\thead{$\downarrow$MSE (ml)} &\thead{$\uparrow$CP$\pm10\%$ (\%)}& \thead{$\uparrow$CP$\pm20\%$ (\%)}  \\
    \midrule
    \makecell{Best estimate}&\makecell{$0.85$}& \makecell{$1.04$}&\makecell{$31.31$}&\makecell{$61.84$}\\ 
    \makecell{Upper lower bound}&\makecell{$0.61$} &\makecell{$0.63$} & \makecell{$52.27$}&\makecell{$75.68$}\\ 
    \bottomrule
    \end{tabular}
    \caption{\textbf{Performance of tidal volume measurement.} Total number of subjects: $3$. Arrows indicate whether higher or lower values signify increased accuracy.  The system demonstrates good agreement with the ventilator's best estimate. This agreement is notably improved when accounting for potential inaccuracies in the ventilator measurements.}
    \label{tab:tv_table}
\end{table}

\begin{table}[H]
    \centering
    \begin{tabular}{c c c c c}
    \toprule
        \thead{Method}& \thead{$\downarrow$MAE (bpm)} &\thead{$\downarrow$MSE (bpm)}&\thead{$\uparrow$CP$\pm5\%$ (\%)} & \thead{$\uparrow$ CP$\pm10\%$ (\%)} \\
    \midrule
        \makecell{CHROM Peaks} & \makecell{$30.90$} & \makecell{$1052.39$} & \makecell{$2.70$}& \makecell{$7.88$}\\
        \makecell{CHROM Fourier} & \makecell{$7.72$} & \makecell{$82.49$} & \makecell{$50.87$} & \makecell{$94.47$}  \\
        \makecell{POS  Peaks} & \makecell{$29.10$} & \makecell{$942.54$} & \makecell{$1.31$} & \makecell{$9.70$} \\
        \makecell{POS Fourier} & \makecell{$7.83$}& \makecell{$85.62$} & \makecell{$51.41$} &\makecell{$91.12$}\\
        \makecell{Combined}& \makecell{$7.69$} & \makecell{$81.24$} & \makecell{$50.84$} & \makecell{$93.37$} \\
    \bottomrule
    \end{tabular}
    \caption{\textbf{Comparison of heart rate estimation methods.} Total number of subjects: $11$. Arrows indicate whether higher or lower values signify increased accuracy.  Peak counting is highly inaccurate due to the noisy signals. Highest performance is achieved by combining the Fourier CHROM and POS signals.}
    \label{tab:heart_rate_table}
\end{table}

\begin{table}[H]
    \centering
    \begin{tabular}{c c c c c}
    \toprule
       \thead{ Method}& \thead{$\downarrow$MAE (\%)} &\thead{$\downarrow$MSE (\%)} & \thead{$\uparrow$CP$\pm3\%$ (\%)}& \thead{$\uparrow$CP$\pm6\%$ (\%)} \\
    \midrule
        \makecell{Red/Infrared} & \makecell{$3.37$}&\makecell{$15.89$}& \makecell{$57.45$} & \makecell{$84.15$}\\
        \makecell{Red/Blue} & \makecell{$8.93$} & \makecell{$123.11$} & \makecell{$20.45$} & \makecell{$41.14$}  \\
        \makecell{YCgCr} &\makecell{$3.27$}& \makecell{$22.53$}& \makecell{$64.23$}& \makecell{$74.23$} \\
        \makecell{Calibration free} & \makecell{$9.49$} & \makecell{$190.34$}& \makecell{$31.65$}& \makecell{$47.13$}\\
    \bottomrule
    \end{tabular}
    \caption{\textbf{Comparison of oxygen saturation estimation methods.} Total number of subjects: 7. Arrows indicate whether higher or lower values signify increased accuracy.  The Red/Infrared ratio of ratio method achieved the highest performance.}
    \label{tab:ox_sat_tab}
\end{table}

\begin{table}[H]
    \centering
    \begin{tabular}{c c c  }
    \toprule
        \thead{Symbol} & \thead{Name} & \thead{Formula}  \\
        \midrule
        \vspace{5pt}
        
        \makecell{MAE} & \makecell{Mean Absolute Error} & \makecell{$\frac{1}{N} \sum_{n=1}^{N} |x_n-y_n|$}\\

        \makecell{MSE} & \makecell{Mean Square Error} & \makecell{$\frac{1}{N} \sum_{n=1}^{N} (x_n-y_n)^2$}\\

        \makecell{CP} & \makecell{Coverage Probability} & \makecell{$P(x_{lower}\leq y \leq x_{upper} ) $}\\
      
         \bottomrule
    \end{tabular}
    \caption{\textbf{Statistical methods.} Definitions of the statistical methods used for comparing two signals.}
    \label{tab:statisticalmetrics}
\end{table}

\section{Data availability}
The datasets generated and/or analysed during the current study are available in the Apollo repository, \url{https://doi.org/10.17863/CAM.111417 }. Raw videos are not publicly available to protect the privacy of neonates involved in the study.

\section{Code availability}
The underlying code for this study is available on GitHub and can be accessed via this link \url{https://github.com/Sr933/Meerkat-Vital-Sign-Analysis-Pipeline.git}. 
\section{Author contributions}
K.B., J.L. and A.G. came up with the idea for the project. S.R.E., A.G., and J.W. performed and contributed to the image/video analysis and signal processing methods. S.R.E. drafted the first version of the manuscript. A.G. and L.T. collected the datasets.  K.B. provided the clinical guidance. J.L. provided the overall guidance for the project. All authors reviewed the manuscript.
\section{Acknowledgements}
This study was funded by the Rosetrees Trust, Isaac Newton Trust and Stoneygate Trust. The funder played no role in study design, data collection, analysis and interpretation of data, or the writing of this manuscript. 
\section{Ethics declaration}
All authors declare no financial or non-financial competing interests.  
\printbibliography

\end{document}